\documentclass[twocolumn,showpacs,preprintnumbers,amsmath,amssymb]{revtex4}
\usepackage{dcolumn}
\usepackage{bm}
\usepackage{latexsym}
\usepackage{amsfonts}
\usepackage[dvips]{graphicx}
\usepackage[usenames]{color}
\usepackage{makeidx}

\newcommand{\pa}{\partial}

\newcommand{\ket}{\rangle }
\newcommand{\bra}{\langle }

\newcommand{\ve}{\varepsilon}
\newcommand{\dket}{\ket\ket}
\newcommand{\dbra}{\bra\bra}

\newcommand{\Vect}[1]{\mbox{\boldmath$#1$}}

\begin{document}
\title{Photovoltaic Hall effect in graphene
}
\author{Takashi Oka and Hideo Aoki}
\address{Department of Physics, University of Tokyo, Hongo, Tokyo 113-0033, 
Japan}

\date{\today}
\begin{abstract}
\noindent
Response of electronic systems in intense lights (AC electric fields) 
to DC source-drain fields is formulated with the Floquet method.  
We have then applied the formalism 
to graphene, for which we show that a non-linear effect of 
a circularly polarized light can open a gap 
in the Dirac cone, which is predicted to lead to a photo-induced dc Hall current.
This is numerically confirmed for a graphene ribbon attached to electrodes 
with the Keldysh Green's function. 

\end{abstract}
\maketitle
{\it Introduction  ---} 
Non-linear phenomena in electronic systems are fascinating 
since they can lead to transport properties qualitatively 
distinct from those in equilibrium. 
In this Letter we seek such a possibility 
by combining (i) the geometric phase 
argument extended to electron transports in 
intense AC fields with (ii) physics of graphene 
involving chiral states associated with two Dirac cones.  
Indeed, the geometric phase has become an important ingredient 
in the modern theory of electric transport\cite{Oka2005a},
which goes back to Thouless's idea of charge pumping 
where he showed that an adiabatic deformation 
of the system may lead to quantized transport\cite{Thouless1983}.
Extensions along several directions have been done subsequently. 
Thouless {\it et al.} (TKNN)\cite{TKNN}
have shown that the Kubo formula for the 
Hall conductivity can be expressed as a topological density.
Berry then showed that such phases are present in general quantum systems, 
and the topological density in the TKNN formula
is now called the Berry curvature \cite{Berry1984}. 
Aharonov and Anandan further extended the notion of 
geometric phases to non-adiabatic situations, i.e.,
Aharonov-Anandan (AA) phase\cite{Aharonov1987}. 

\begin{figure}[t]
\centering 
\includegraphics[width=8.75cm]{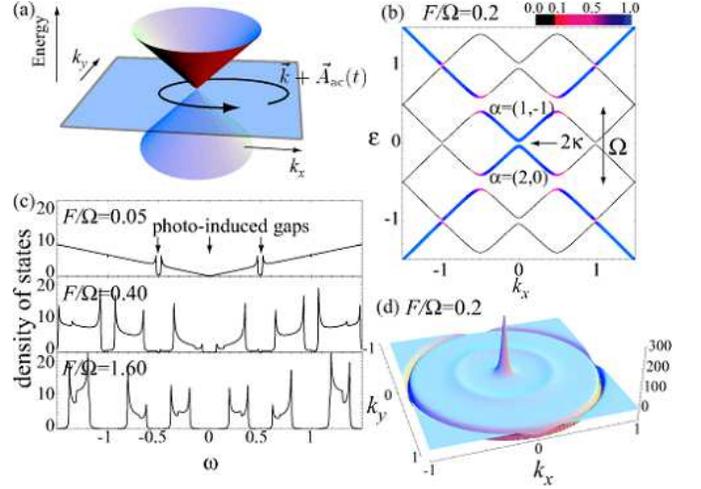}
\caption{(a) A trajectory of $\Vect{k}+\Vect{A}_{\rm ac}(t)$ around 
a Dirac point in a circularly polarizated light field. 
(b) The Floquet quasi-energy (black curves) plotted against $k_x$ 
with $k_y=0$ for $F/\Omega =0.2$. The color coding
represents the weight of the static ($m=0$) component. 
(c) Density of states for various field strengths. 
(d) The photo-induced Berry curvature $\left[\nabla_{\Vect{k}}\times\Vect{\mathcal{A}}_\alpha(\Vect{k})\right]_z$ for $\alpha=(1,m)$ 
for $F/\Omega =0.2$. 
The series $\alpha=(2,m)$ have a similar behavior with an inverted sign. 
Frequency is $\Omega/v=1$ throughout.
}
\label{fig:dirac}
\end{figure}

With these as a background, the concept of the present 
study, with graphene in mind, is simple. 
Let us put a crystal with a Dirac band
in a {\it circularly polarized light}.
As in Fig. \ref{fig:dirac}(a), 
an intense AC field $\Vect{A}_{\rm ac}(t)$ will 
deform the single-body Hamiltonian, 
and each $k$-point will follow a circle in the Brilliouin zone. 
If the loop encircles the Dirac point, 
non-adiabatic charge pumping should take place, in which 
the wave function acquires a non-trivial AA phase. 
So the question is: can this be detected with a DC transport measurement?  
In order to answer this, 
we have formulated the Kubo formula for the DC response 
for systems in intense AC fields, which is accomplished with 
the Floquet-matrix formalism 
by which we can solve the 
time-dependent dynamics of $k$-points within a static approach.
We shall show that a 
TKNN-like formula for the Hall conductivity is 
obtained, where the Berry curvature 
is now expressed in terms of Floquet states which depends on the AA phase. 
We apply this formula to a single Dirac band first, 
then a graphene. 
For graphene\cite{Novoselov07262005,Novoselov10222004}, 
which is dominated by the chirality, 
we conclude that a photo-induced Hall current ({\it despite the 
absence of uniform magnetic fields}) should appear in graphene 
irradiated by circularly polarized light and attached to two electrodes, 
where the Hall current can exceed the longitudinal current in magnitude.

{\it Kubo formula in the presence of strong light fields  ---} 
We first derive the Kubo formula for electric transport 
for systems in strong AC fields, where we concentrate on 
the one-body problem for simplicity.  
The AC electric field is introduced 
as a time-dependent gauge potential $\Vect{A}_{\rm ac}(t)$, 
which satisfies $\Vect{A}_{\rm ac}(t+T)=
\Vect{A}_{\rm ac}(t)$ with $T$ the periodicity 
(while the frequency is $\Omega=2\pi/T$).  
On top of this we introduce, as in the linear response, 
a weak gauge potential $\Vect{A}(t)=\Vect{E}t$ 
that changes slowly to represent an infinitesimal DC electric field $E$, 
where we have set $e=1,\;\hbar =1$.  
Thus we have a time-dependent Hamiltonian, 
\begin{eqnarray}
H(t)=\int \frac{d\Vect{k}}{(2\pi)^d}\Psi^\dagger(\Vect{k})
h\left(\Vect{k}+\Vect{A}_{\rm ac}(t)
+\Vect{A}(t) \right)\Psi(\Vect{k}),
\end{eqnarray}
where $h(\Vect{k})$ is the one-body Hamiltonian 
and $\Psi$ a state vector (which is multi-dimensional 
for multi-bands).  
To represent the states in AC fields, we can 
employ the Floquet operator (see \cite{HanggiREVIEW1998,KohlerREVIEW2005}) 
$
\mathcal{H}(\Vect{k},\Vect{A}_{\rm ac}(t),\Vect{A}(t))=h\left(\Vect{k}+\Vect{A}_{\rm ac}(t)
+\Vect{A}(t) \right)-i\pa_t,
$
with which the time-dependent Schr\"odinger equation reads 
$
\mathcal{H}(\Vect{k},\Vect{A}_{\rm ac}(t),\Vect{A}(t))|\Psi(\Vect{k};t)\ket=0.
$
Since $\Vect{A}(t)$ is infinitesimal and adiabatically 
changing, we take it as the adiabatic parameter, 
while the AC field is intense and rapidly oscillating.  So, 
for each interval of time over which $\Vect{A}$ may be considered 
to be constant, we can 
introduce the Floquet states (a time-analog of Bloch states) 
satisfying the Floquet equation $
\mathcal{H}(\Vect{k},\Vect{A}_{\rm ac}(t),\Vect{A})|\Phi_\alpha(\Vect{k};\Vect{A},t)\ket=\ve_\alpha(\Vect{k};\Vect{A})|\Phi_\alpha(\Vect{k};\Vect{A},t)\ket
$
with a 
periodicity $|\Phi_\alpha(\Vect{k};\Vect{A},t+T)\ket=|\Phi_\alpha(\Vect{k};\Vect{A},t)\ket$ 
\cite{PhysRevA.7.2203},
where $\ve_\alpha(\Vect{k};\Vect{A})$ is called the Floquet quasi-energy 
which is a sum of the dynamical phase and the AA phase (see eqn.(\ref{eq:quasienergysum}) below), 
and $\alpha$ labels the eigenstate.  
The solution of the 
time-dependent Schr\"odinger equation 
for a fixed $\Vect{A}$ can be expressed as
$|\Psi_\alpha(t)\ket=e^{-i\ve_\alpha t}|\Phi_\alpha(t)\ket$. 
If we define an inner product averaged over a period by
$\dbra\alpha|\beta\dket\equiv \frac{1}{T}\int_0^T dt\bra \alpha(t)|
\beta(t)\ket$, the Floquet states form 
an orthonormal basis, i.e., 
$
\dbra \Phi_{\alpha }(\Vect{k};\Vect{A})|\Phi_{\beta}(\Vect{k};\Vect{A})\dket
=\delta_{\alpha\beta}.$
With these as a basis 
the solution to the time-dependent Schr\"odinger equation 
for the slow change in $\Vect{A}$ is
\begin{eqnarray}
&&|\Psi(\Vect{k};\Vect{A}(t),t)\ket=
e^{-i\int_0^t dt'\ve_\alpha({\small \Vect{k}};{\small \Vect{A}}(t'))}\left[
|\Phi_\alpha(\Vect{k};\Vect{A}(t),t)\ket+\right.\nonumber
\\
&&\left.\sum_{\beta\ne\alpha}
|\Phi_\beta(\Vect{k};\Vec{A}(t),t)\ket\frac{\dbra \Phi_\beta(\Vect{k};\Vect{A}(t))|
\frac{\pa{\small \Vect{A}}}{\pa t}\cdot\frac{\pa}{\pa {\small \Vect{A}}}|\Phi_\alpha(\Vect{k};\Vect{A}(t))\dket}
{\ve_{\beta}(\Vect{k};\Vect{A}(t))-\ve_{\alpha}(\Vect{k};\Vect{A}(t))}
\right],\nonumber
\end{eqnarray}
up to first order in time derivatives,
with the $\alpha$-th Floquet state taken to be the initial state. 
One can readily derive this result with the 
two-time method\cite{peskin:4590,Breuer1989}. 
We can immediately notice that the geometrical phase appears 
in a form $
\dbra \Phi_\beta (\Vect{k};\Vect{A}(t))|\frac{\pa{\small \Vect{A}}}{\pa t}
\cdot\frac{\pa}{\pa {\small \Vect{A}}}|\Phi_\alpha (\Vect{k};\Vect{A}(t))\dket
= \sum_{\beta \neq \alpha} \dbra\Phi_\beta (\Vect{k};\Vect{A}(t))|
\frac{\pa{\small \Vect{A}}}{\pa t}\cdot\frac{\pa \mathcal{H}}{\pa {\small \Vect{A}}}
|\Phi_\alpha (\Vect{k};\Vect{A}(t))\dket[\ve_\beta(\Vect{A}(t))-\ve_\alpha(\Vect{A}(t))]^{-1}$.  Since the current operator is 
$\Vect{J} = \pa h(\Vect{k}+\Vect{A}_{\rm ac}(t)+\Vect{A})/\pa \Vect{A}$, 
the above formula for the DC transport in 
an intense AC background field is rewritten as
\begin{eqnarray}
&&\sigma_{ab}(\Vect{A}_{ac})=i\int \frac{d\Vect{k}}{(2\pi)^d}
\sum_{\alpha,\beta\ne\alpha}
\frac{[f_\beta(\Vect{k})-f_\alpha(\Vect{k})]}{\ve_\beta(\Vect{k})-\ve_\alpha(\Vect{k})}\nonumber
\\
&&\hspace{0.8cm}\times
\frac{
\dbra\Phi_\alpha(\Vect{k})|J_b|\Phi_{\beta}(\Vect{k})\dket
\dbra\Phi_\beta(\Vect{k})|J_a|\Phi_{\alpha}(\Vect{k})\dket
}{\ve_\beta(\Vect{k})-\ve_\alpha(\Vect{k})+i\eta},
\end{eqnarray}
where $f_\alpha(\Vect{k})$ is the non-equilibrium distribution 
(occupation fraction) of the $\alpha$-th Floquet state, 
$\eta$ a positive infinitesimal, 
and we have put the perturbation $\Vect{A}=0$ as in the 
linear-response theory. 
The essential difference from the conventional 
Kubo formula in the absence of AC fields is that 
the energy is replaced with the Floquet quasi-energy, and 
the inner product with a time averaged one.  
We note that similar expressions were obtained by
Torres and Kunold in their study of 
microwave-assisted zero-resistance states\cite{torres:115313}.
The Hall conductivity can be further simplified 
to a TKNN-like formula, 
\begin{eqnarray}
\sigma_{xy}(\Vect{A}_{\rm ac})=e^2\int \frac{d\Vect{k}}{(2\pi)^d}\sum_\alpha
f_\alpha(\Vect{k})\left[\nabla_{\Vect{k}}\times\Vect{\mathcal{A}}_\alpha(\Vect{k})\right]_z ,
\end{eqnarray}
where 
$\Vect{\mathcal{A}}_\alpha(\Vect{k}) \equiv 
-i\dbra\Phi_\alpha(\Vect{k})|\nabla_{\Vect{k}}|\Phi_\alpha(\Vect{k})\dket$.
We note that if we separate the Floquet index into  $\alpha=(i,m)$
where $i$ labels the original band and $m$ is the Floquet index, then 
$\Vect{\mathcal{A}}_\alpha(\Vect{k})$
is independent of $m$. 
However, the occupation $f_\alpha(\Vect{k})$ depends on both indices.
In equilibrium, $f_{(i,m)}(\Vect{k})=\delta_{m0}f_{\rm FD}(E_i(\Vect{k}))$
holds where $E_i(\Vect{k})$ is the energy of $i$th state 
and $f_{\rm FD}$ the Fermi-Dirac distribution. 
In non-equilibrium, however, the distribution is {\it non-universal}, 
and depends on the detail of the system such as how the 
electrodes are attached, etc., 
so that a case-by-case study should be needed to 
determine the distribution and hence the DC transport of the system.

{\it Application to a Dirac band ---}
Now we study the effect of AC fields on the two-dimensional Dirac band. 
The Hamiltonian is 
$H(t)=\tau_zv[k^x+A_{\rm ac}^x(t)]\sigma_x+v[k^y+A_{\rm ac}^y(t)]\sigma_y$,
where $\tau_z=\pm 1$ labels the chirality ($K$ and $K'$ points in graphene), 
$v$ is the velocity (set to $v=1$ hereafter), 
and  $\sigma_i$ the Pauli matrices.
The circularly polarized field is given as
$(A_{\rm ac}^x, A_{\rm ac}^y) = A(\cos\Omega t, \sin\Omega t)$ 
where $A \equiv F/\Omega$ with $F$ being the field strength. 
Here we neglect the momentum of light since $v\ll c$,
and consider direct transitions, 
which is a situation different from the Volkov solution\cite{Volkov35}. 
With Fourier transformed Floquet states $|\Phi(t)\ket=
\sum_me^{-im\Omega t}|u_\alpha^m\ket$ the Floquet equation becomes 
\begin{eqnarray}
\sum_{n}H^{mn}|u^n_\alpha\ket=(\ve_\alpha+m\Omega)|u^m_\alpha\ket
,
\end{eqnarray}
where the Floquet Hamiltonian 
$H^{mn}=\frac{1}{T}\int_0^TdtH(t)e^{i(m-n)\Omega t}$
has 
$H^{mm}=\left(^{0\;k_x-ik_y}_{k_x+ik_y\;0}\right)$
for the diagonal components,
whereas the off-diagonal components 
depend on $\tau_z$, i.e., $H^{mm+1}=\left(^{0A}_{00} \right),
\;H^{mm-1}=\left(^{00}_{A0} \right)$
for $\tau_z=1$
and $H^{mm+1}=\left(^{00}_{-A0} \right),
\;H^{mm-1}=\left(^{0-A}_{00} \right)$
for $\tau_z=-1$.
This defines an eigenvalue problem for a block tri-diagonal matrix
that can be solved numerically with a truncation at certain $|m|$.

As an important effect of the AC field, 
gaps open at $\omega=$ integer $\times \Omega/2$ 
in the quasi-energy band structure (Fig.1(b)), 
reflecting a Dirac-band analog of the 
AC-Wannier-Stark ladder, 
as also seen in the density of states 
$A(\omega)=-\frac{1}{\pi}\int d\Vect{k}\sum_\alpha\mbox{Im}
\frac{\bra u^0_\alpha|u^0_\alpha\ket}{\omega-\ve_\alpha+i\eta}$ 
in  Fig. \ref{fig:dirac}(c).  
The gap at $\omega=\pm \Omega/2$ is the largest, 
which is related to one-photon assisted transport discussed later.
We note that this gap also opens in the linearly polarized case as 
studied in \cite{syzranov:045407}.
More importantly, a new gap opens at $\Vect{k}=0,\;\ve=0$ in the circularly polarized case. 
Indeed, one can show that the solution of the 
time-dependent Schr\"odinger equation 
at $\Vect{k}=0$ is  
\begin{eqnarray}
|\Psi_\alpha(\Vect{k}=0,t)\ket \propto e^{-i\ve_\alpha t}\left(
	\begin{array}{c}
1\\
\frac{\tilde{\ve}_i}{A}e^{i\Omega t}
\end{array}
\right)
\end{eqnarray}
where the Floquet state is labeled by 
$\alpha=(i,m)$ with $i=1,2$ representing the
upper and lower branches of the Dirac band, 
$m$ the Floquet index. 
The quasi-energy is $
\ve_\alpha=\tilde{\ve}_i+m\Omega$
with  $\tilde{\ve}_1=
\frac{\sqrt{4A^2+\Omega^2}+\Omega}{2},\;
\tilde{\ve}_2=
\frac{-\sqrt{4A^2+\Omega^2}+\Omega}{2}$.
The $\alpha=(1,-1),\;(2,0)$ bands are direct descendants 
of the original Dirac bands, and the dynamical gap $2\kappa$ opening between 
them is $
\kappa=\ve_{(1,-1)}=
\frac{\sqrt{4A^2+\Omega^2}-\Omega}{2}$.
The dynamical gap first grows quadratically with $A$, 
$2\kappa\sim 2A^2/\Omega$, followed by an asymptote 
$2\kappa\sim 2A-\Omega$.  
An important property of the quasi-energy is that 
it is a sum of the dynamical phase 
and the AA phase, i.e.,
\begin{eqnarray}
\ve_\alpha=\dbra \Phi_\alpha|H(t)|\Phi_\alpha\dket+\gamma_\alpha^{\rm AA}/T,
\label{eq:quasienergysum}
\end{eqnarray}
where the AA phase is given by 
\begin{eqnarray}
\gamma^{\rm AA}_\alpha \equiv T\dbra \Phi_\alpha|i\pa_t|\Phi_\alpha\dket = \pm\pi\left\{[4(A/\Omega)^2+1]^{-1/2}-1\right\},
\end{eqnarray}
where $\pm$ refers to $\alpha=(1,-1),(2,0)$. 
In the adiabatic limit ($\Omega\to 0$ 
with a fixed $A$) it approaches to $\mp \pi$. 
We note that only the $k$-points with $|\Vect{k}|<A$ acquires the 
AA phase, since otherwise the Dirac cone is not encircled.  
In the Berry curvature for 
the $\alpha=(2,0)$ Floquet state (Fig. \ref{fig:dirac}(d)) 
there is a conspicuous peak around $\Vect{k}= 0$,
\begin{eqnarray}
\left[\nabla_k\times\Vect{\mathcal{A}}_\alpha(\Vect{k})\right]_z \sim 
\pm\tau_z\frac{1}{2}\kappa(|\Vect{k}|^2+\kappa^2)^{-3/2},
\label{eq:Berrycurvature}
\end{eqnarray}
where $\pm$ corresponds to $\alpha=(1,m),(2,m)$.  
In this expression, two geometric quantities appear, 
where the Berry curvature comes from the perturbative 
treatment of the weak DC electric field whereas the AA 
phase emerges because of the time-periodic dynamics of 
$k$-points in intense AC fields.  
Due to the factor $\tau_z$, 
the contribution from K and K' points in 
graphene will cancel with each other 
if the distribution is identical between them.  
However, we shall see that, if we 
apply a static DC bias across the system, 
chirality (the valley symmetry) is degraded, which 
will lead to a non-trivial curvature, hence to a Hall current.

{\it Keldysh approach to photovoltaic transport in graphene ---} 
So we move on to a Keldysh 
Green's function analysis of transport properties in a graphene 
irradiated by a circularly polarized light and 
attached to two electrodes.  
The system is described by an action, 
$
S = \int_Cdt\left(
\mathcal{L}_{\rm graphene}+\mathcal{L}_{\rm mix}
+\mathcal{L}_{\rm electrodes}\right),$
where $
\mathcal{L}_{\rm graphene}=
\sum_{i\ne j}c^\dagger_i(i\pa_t
-t_{ij}e^{iA_{ij}^{\rm ac}(t)}) c_j$ is the tight-binding 
model for graphene with a hopping $t_{ij}=-w$ for nearest neighbors, 
while $
\mathcal{L}_{\rm mix}=\sum_{k,r}V^r_{\rm mix}\left[(a_{k}^r)^\dagger c_r
+\mbox{h.c.}\right]$ represents the coupling between the 
electrodes and graphene, with the spin degrees of freedom ignored.  
The AC field is introduced by
$A_{ij}^{\rm ac}(t)=(\Vect{r}_i-\Vect{r}_j)\cdot\Vect{A}^{\rm ac}(t)$
with $\Vect{A}^{\rm ac}(t)=(F/\Omega)(\cos\Omega t, \sin\Omega t)$, 
where $F=eaE$ is the normalized field strength ($a$: lattice const.).  
We assume that the electrodes are described by a 
fermion operator $a^r$ 
($r\in\{\rm L,R\}$ labeling the left and right electrodes), 
for which a Fermi-Dirac distribution 
$\bra a^{r^\dagger} a^r\ket=f_r = [e^{\beta(\omega-\mu_r)}+1]^{-1}$ 
is assumed with electrode-dependent chemical potential $\mu_r$.
They are related to the DC bias $V$ across the electrodes 
by $\mu_L=V/2,\;\mu_R=-V/2$.   
Imposing a periodic boundary condition in the direction ($y$) of the 
graphene ribbon, 
the Keldysh Green's functions for each momentum $k_y$ 
become a matrix labeled by the site in the $x$-direction  ($i=1,\ldots,N$)
and by the Floquet index.
The Green's functions satisfy
\begin{widetext}
\begin{eqnarray}
&&\left(
	\begin{array}{cc}
G^{R}_{k_y}(\omega)&
G^{K}_{k_y}(\omega)\\
0&G^{A}_{k_y}(\omega)
\end{array}
\right)^{-1}_{ij;mn}
=\left(
	\begin{array}{cc}
(\omega+n\Omega+i\eta)\delta_{mn}\delta_{ij}-(\hat{H})^{mn}_{ij}(k_y)
&0\\
0&(\omega+n\Omega-i\eta)\delta_{mn}\delta_{ij}-(\hat{H})^{mn}_{ij}(k_y)
\end{array}
\right)\nonumber\\
&&+\delta_{i1}\delta_{mn}\left(
	\begin{array}{cc}
i\Gamma_L/2&-i\Gamma_L(1-2f_L(\omega+m\Omega))\\
0&-i\Gamma_L/2
\end{array}
\right)+\delta_{iN}\delta_{mn}\left(
	\begin{array}{cc}
i\Gamma_R/2&-i\Gamma_R(1-2f_R(\omega+m\Omega))\\
0&-i\Gamma_R/2
\end{array}
\right),\nonumber
\end{eqnarray}
\end{widetext}
where $G^{K,R,A}$ are the Keldysh, retarded, 
and advanced Green's function, respectively, 
$\Gamma_r \propto |V^r_{\rm mix}|^2$ is the imaginary part of the self-energy 
due to the sample-electrode coupling.   
The effective Floquet Hamiltonian is 
defined by $(\hat{H})^{mn}(k_y)=\frac{1}{T}\int_{0}^{T}dt
e^{i(m-n)\Omega t}\hat{H}(k_y;A^{\rm ac}(t))$, where 
the indices $i,\;j$ are suppressed. 
The current between sites $i,j$ is determined from the lesser 
component $G^<$ by
\begin{eqnarray}
\bra J^a_{ij}(t)\ket 
= -i\frac{ e}{\hbar}\frac{1}{N_{k_y}}\sum_{k_y}
\sum_{mn}\int_{0}^{\Omega}
\frac{d\omega}{2\pi}e^{-i(m-n)\Omega t}
\\
\times
\left[J_{ij}(t)
(G^<_{k_y})_{mn;ji}(\omega)+J_{ji}(t)
(G^<_{k_y})_{mn;ij}(\omega)\right],\nonumber
\end{eqnarray}
where $J_{ij} = \delta \hat{H}/\delta A_{ij}$ is the current operator. 
In practice, we do not calculate the
Keldysh component but use the Keldysh equation (c.f. \cite{JauhoWingreenMeir94})
to relate $G^<$ 
with $G^{A}, G^{R}$ 
obtained by diagonalizing the
Floquet Hamiltonian.

\begin{figure}[t]
\centering 
\includegraphics[width=8.5cm]{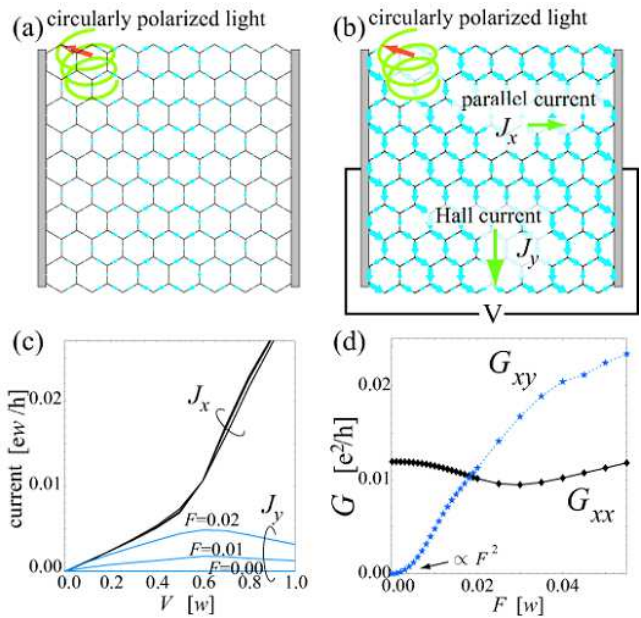}
\caption{
DC current distribution in an armchair 
graphene ribbon attached to 
electrodes subject to 
a circularly polarized light for a finite AC 
field $F=0.025w, \Omega=0.3w$ with no DC bias (a) and
with a finite bias $V=0.005w$ (b).  
(c) $I-V$ characteristics of the
longitudinal $J_x$ (black) and the Hall current $J_y$ (blue) 
for various values of $F$. 
(d) DC conductance $G=J/V$ plotted against field strength $F$ 
for a fixed bias $V=0.005w$. 
System size in the $x$-direction $N=34$ throughout.
}
\label{fig:current}
\end{figure}

\begin{figure}[t]
\centering 
\includegraphics[width=8.cm]{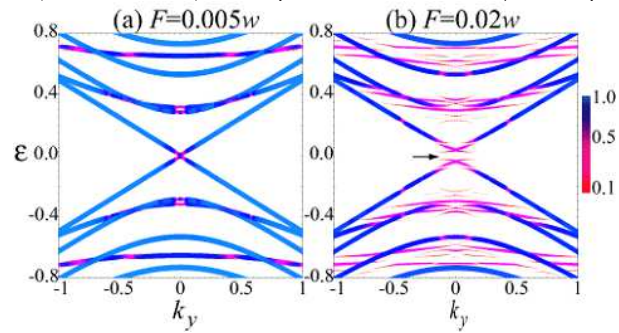}
\caption{
Floquet quasi-energy diagram for an armchair\cite{zigzag} 
ribbon in an AC field 
for $F=0.005w$(a), $0.02w$(b)
with frequency $\Omega=0.3w$. 
The color coding represents 
the static weight $|\bra u_\alpha^0|u^0_\alpha\ket|^2$ as 
in Fig. \ref{fig:dirac}(b).
}
\label{fig:spec}
\end{figure}

In the obtained current distribution (Fig.\ref{fig:current}(a)) 
the polarized light induces locally circulating currents in the 
absence of the bias $V$ across the electrodes. 
There is no net current in the $y$-direction as it should. 
This current resembles the orbital magnetism 
which was predicted to arise when a
perturbation induce a gap in the Dirac cone 
\cite{xiao:137204,thonhauser:137205}.
Indeed, the circularly polarized light in the present case 
opens a gap around the band crossing 
(arrow in Fig.\ref{fig:spec}(b)), 
which has led to a similar DC effect. 

A striking finding here is that, 
when we switch on the bias voltage $V$, 
we have a {\it photo-induced DC Hall current} as well as 
a longitudinal ($\parallel x$) current (Fig.~\ref{fig:current}~(b)).  
The Hall current is natrually 
inverted when the right circularly polarized field 
$\Vect{A}_{\rm ac}\propto(\cos\Omega t, \sin\Omega t)$ is changed 
into the left polarization $\propto(\cos\Omega t, -\sin\Omega t)$, 
or the bias voltage is inverted.  
The $I-V$-characteristics is shown 
in Fig.~\ref{fig:current}~(c) 
for $J_x$ and $J_y$, the averaged current 
in $x$ and $y$ directions, respectively. 
The photo-induced net Hall current 
grows linearly with the bias $V$, but saturates and then decrease 
when $V$ becomes large.
Figure\ref{fig:current}(d) depicts the dependence of 
the conductance, $G_{xx}=J_x/V,\;G_{xy}=J_y/V$ on the intensity $F$ 
of the circularly polarized light for a fixed bias $V$.
The conductance, unlike the conductivity discussed in the first part 
of the paper, is less universal and depends on the contact 
$\Gamma_r$ (here we set $\Gamma_L=\Gamma_R=0.2w$) etc, 
but we expect that the effect will be 
qualitatively robust.  
As we increase $F$, the longitudinal $G_{xx}$ first decreases, and 
increase again ($F>0.03w$).
The decrease can be explained by the 
gap opening at the Dirac points, 
while the increase is due to photo-assisted transport.  
We note that similar features have been 
experimentally observed in microwave irradiated carbon nanotubes \cite{PhysRevB.70.153402}. 
The Hall conductance $G_{xy}$, on the other hand, 
initially grows quadratically with $F$ 
and then increases linearly, 
which is a dependence similar to the gap $\kappa$ in 
the Dirac cone (eqn.(\ref{eq:Berrycurvature})), 
which may indicate $J_y\sim \kappa V$.  
We note that 
a similar expression was obtained in the case where the chirality is 
broken in a static manner\cite{xiao:236809}.



To summarize, we have found that 
a combined effect of an intense AC field of a 
circularly polarized light and a (weak) DC bias 
can produce a photo-voltaic dc Hall current in graphene, 
despite the absence of a uniform magnetic field.  
The typical intensity of laser conceived here, 
$F\sim 0.001w$, corresponds to 
$E\sim 10^7\mbox{V/m}$ 
for photon energy $\Omega\sim 1\mbox{eV}$, $w=2.7\mbox{eV},\;a=2.6\AA$, 
which should be within the experimental feasibility.  
Inclusion of dissipation etc will be an interesting future problem.

We wish to thank Andre Geim for fruitful discussion in the
initial stage of this work.
TO thanks Jun Okubo and Naoto Tsuji for eliminating discussions
on the Floquet method. 
HA was supported by a Grant-in-Aid for Scientific Research on Priority Area
``Anomalous quantum materials" from the Japanese 
Ministry of Education, TO by Grant-in-Aid for young Scientists (B).


\begin{thebibliography}{18}
\expandafter\ifx\csname natexlab\endcsname\relax\def\natexlab#1{#1}\fi
\expandafter\ifx\csname bibnamefont\endcsname\relax
  \def\bibnamefont#1{#1}\fi
\expandafter\ifx\csname bibfnamefont\endcsname\relax
  \def\bibfnamefont#1{#1}\fi
\expandafter\ifx\csname citenamefont\endcsname\relax
  \def\citenamefont#1{#1}\fi
\expandafter\ifx\csname url\endcsname\relax
  \def\url#1{\texttt{#1}}\fi
\expandafter\ifx\csname urlprefix\endcsname\relax\def\urlprefix{URL }\fi
\providecommand{\bibinfo}[2]{#2}
\providecommand{\eprint}[2][]{\url{#2}}


\bibitem{Oka2005a} Geometric phase in nonequilibrium 
is evoked in e.g. the dielectric 
breakdown of a Mott insulator in 
T. Oka and H. Aoki, Phys. Rev. Lett. {\bf 95}, 137601 (2005).

\bibitem[{\citenamefont{{D. J. Thouless}}(1983)}]{Thouless1983}
\bibinfo{author}{\bibnamefont{{D. J. Thouless}}}, \bibinfo{journal}{Phys. Rev.
  B} \textbf{\bibinfo{volume}{27}}, \bibinfo{pages}{6083}
  (\bibinfo{year}{1983}).

\bibitem[{\citenamefont{{D. J. Thouless, M. Kohmoto, M. P. Nightingale and M.
  den Nijs}}(1982)}]{TKNN}
\bibinfo{author}{\bibnamefont{{D. J. Thouless, {\it et al.}}}}, \bibinfo{journal}{Phys. Rev. Lett.}
  \textbf{\bibinfo{volume}{49}}, \bibinfo{pages}{405} (\bibinfo{year}{1982}).

\bibitem[{\citenamefont{{M. V. Berry}}(1984)}]{Berry1984}
\bibinfo{author}{\bibnamefont{{M. V. Berry}}}, \bibinfo{journal}{Proc. R. Soc.
  Lond.} \textbf{\bibinfo{volume}{A 392}}, \bibinfo{pages}{45}
  (\bibinfo{year}{1984}).

\bibitem[{\citenamefont{{Y. Aharonov and J. Anandan}}(1987)}]{Aharonov1987}
\bibinfo{author}{\bibnamefont{{Y. Aharonov and J. Anandan}}},
  \bibinfo{journal}{Phys. Rev. Lett.} \textbf{\bibinfo{volume}{58}},
  \bibinfo{pages}{1593} (\bibinfo{year}{1987}).

\bibitem[{\citenamefont{{K. S. Novoselov, D. Jiang, F. Schedin, T. J. Booth, V.
  V. Khotkevich, S. V. Morozov and A. K. Geim}}(2005)}]{Novoselov07262005}
\bibinfo{author}{\bibnamefont{{K. S. Novoselov, {\it et al.}}}},
  \bibinfo{journal}{Proc. Nat. Aca. Sci. } \textbf{\bibinfo{volume}{102}},
  \bibinfo{pages}{10451} (\bibinfo{year}{2005}).

\bibitem[{\citenamefont{{K. S. Novoselov, A. K. Geim, S. V. Morozov, D. Jiang,
  Y. Zhang, S. V. Dubonos, I. V. Grigorieva, A. A.
  Firsov}}(2004)}]{Novoselov10222004}
\bibinfo{author}{\bibnamefont{{K. S. Novoselov, {\it et al.}}}},
  \bibinfo{journal}{Science} \textbf{\bibinfo{volume}{306}},
  \bibinfo{pages}{666} (\bibinfo{year}{2004}).

\bibitem[{\citenamefont{{P. H\"anggi}}(1988)}]{HanggiREVIEW1998}
\bibinfo{author}{\bibnamefont{{P. H\"anggi}}}, \emph{\bibinfo{title}{Quantum
  transport and dissipation}} (\bibinfo{publisher}{WILEY-VCH},
  \bibinfo{year}{1988}), chap.~\bibinfo{chapter}{5}.

\bibitem[{\citenamefont{{Kohler}}(2005)}]{KohlerREVIEW2005}
\bibinfo{author}{\bibnamefont{{S. Kohler, J. Lehmann, and P. H\"anggi}}}, \bibinfo{journal}{Physics Reports},
  \textbf{\bibinfo{volume}{406}}, \bibinfo{pages}{379} (\bibinfo{year}{2005}).

\bibitem[{\citenamefont{{H. Sambe}}(1973)}]{PhysRevA.7.2203}
\bibinfo{author}{\bibnamefont{{H. Sambe}}}, \bibinfo{journal}{Phys. Rev. A}
  \textbf{\bibinfo{volume}{7}}, \bibinfo{pages}{2203} (\bibinfo{year}{1973}).

\bibitem[{\citenamefont{{U. Peskin and N. Moiseyev}}(1993)}]{peskin:4590}
\bibinfo{author}{\bibnamefont{{U. Peskin and N. Moiseyev}}},
  \bibinfo{journal}{J. Chem. Phys.}
  \textbf{\bibinfo{volume}{99}}, \bibinfo{pages}{4590} (\bibinfo{year}{1993})

\bibitem[{\citenamefont{{H. P. Breuer and M. Holthaus}}(1989)}]{Breuer1989}
\bibinfo{author}{\bibnamefont{{H. P. Breuer and M. Holthaus}}},
  \bibinfo{journal}{Phys. Lett. A} \textbf{\bibinfo{volume}{140}},
  \bibinfo{pages}{507} (\bibinfo{year}{1989}).

\bibitem[{\citenamefont{{M. Torres and A. Kunold}}(2005)}]{torres:115313}
\bibinfo{author}{\bibnamefont{{M. Torres and A. Kunold}}},
  \bibinfo{journal}{Phy. Rev. B} \textbf{\bibinfo{volume}{71}},
  \bibinfo{eid}{115313} (\bibinfo{year}{2005})

\bibitem[{\citenamefont{{D. M. Volkov}}(1935)}]{Volkov35}
\bibinfo{author}{\bibnamefont{{D. M. Volkov}}}, \bibinfo{journal}{Z. Phys.}
  \textbf{\bibinfo{volume}{94}}, \bibinfo{pages}{250} (\bibinfo{year}{1935}).

\bibitem[{\citenamefont{Syzranov et~al.}(2008)\citenamefont{Syzranov, Fistul,
  and Efetov}}]{syzranov:045407}
\bibinfo{author}{\bibfnamefont{S.~V.} \bibnamefont{Syzranov}},
  \bibinfo{author}{{\it et al.}}, \bibinfo{journal}{Phys. Rev. B} \textbf{\bibinfo{volume}{78}}, \bibinfo{eid}{045407} (\bibinfo{year}{2008}).

\bibitem[{\citenamefont{{A. P. Jauho, N. S. Wingreen and Y.
  Meir}}(1994)}]{JauhoWingreenMeir94}
\bibinfo{author}{\bibnamefont{{A. P. Jauho, {\it et al.}}}},
  \bibinfo{journal}{Phys. Rev. B} \textbf{\bibinfo{volume}{50}},
  \bibinfo{pages}{5528} (\bibinfo{year}{1994}).

\bibitem[{\citenamefont{{Di Xiao, J. Shi and Qian Niu}}(2005)}]{xiao:137204}
\bibinfo{author}{\bibnamefont{{D. Xiao, {\it et al.}}}},
  \bibinfo{journal}{Phys. Rev. Lett.} \textbf{\bibinfo{volume}{95}},
  \bibinfo{eid}{137204}  (\bibinfo{year}{2005})

\bibitem[{\citenamefont{{T. Thonhauser, D. Ceresoli, D. Vanderbilt and R.
  Resta}}(2005)}]{thonhauser:137205}
\bibinfo{author}{\bibnamefont{{T. Thonhauser, {\it et al.}}}}, \bibinfo{journal}{Phys. Rev. Lett.}
  \textbf{\bibinfo{volume}{95}}, \bibinfo{eid}{137205}
  (\bibinfo{year}{2005}).

\bibitem[{\citenamefont{{J. Kim, and H. Mi. So, N. Kim, J. J. Kim and K.
  Kang}}(2004)}]{PhysRevB.70.153402}
\bibinfo{author}{\bibnamefont{{J. Kim, {\it et al.}}}}, \bibinfo{journal}{Phys. Rev. B} \textbf{\bibinfo{volume}{70}},
  \bibinfo{pages}{153402} (\bibinfo{year}{2004}).


\bibitem[{\citenamefont{Xiao et~al.}(2007)\citenamefont{Xiao, Yao, and
  Niu}}]{xiao:236809}
\bibinfo{author}{\bibfnamefont{D.}~\bibnamefont{Xiao}},
  \bibinfo{author}{{\it et al.}},
  \bibinfo{journal}{Phys. Rev. Lett.} \textbf{\bibinfo{volume}{99}},
  \bibinfo{eid}{236809} (\bibinfo{year}{2007}).

\bibitem{zigzag} 
We have also studied zigzag nano-ribbons 
to find that the photovoltaic Hall current 
behaves similarly, and that the zero-energy zigzag edge modes 
do not alter the physics. 

\end{thebibliography}

\end{document}